\documentclass[twocolumn,aps, amsmath,amssymb, superscriptaddress]{revtex4-2}
\usepackage[utf8]{inputenc}
\usepackage{graphicx}
\def\be{\begin{equation}}
\def\ee{\end{equation}}
\def\bea{\begin{eqnarray}}
\def\eea{\end{eqnarray}}

\usepackage{color}
\usepackage{bm}

\begin{document}

\title{Role of crystal field ground state in the classical spin-liquid behavior of a quasi-one dimensional spin-chain system  Sr$_3$NiPtO$_6$}
\author{V.~K.~Anand}
\email{vivekkranand@gmail.com}
\affiliation{ISIS Neutron and Muon Facility, STFC, Rutherford Appleton Laboratory, Chilton, Oxfordshire, OX11 0QX, UK}
\affiliation{Department of Physics, University of Petroleum and Energy Studies, Dehradun, Uttarakhand, 248007, India}
\affiliation{Department of Mathematics and Physics, University of Stavanger, 4036 Stavanger, Norway}
\author{D.~T.~Adroja }
\email{devashibhai.adroja@stfc.ac.uk}
\affiliation{ISIS Neutron and Muon Facility, STFC, Rutherford Appleton Laboratory, Chilton, Oxfordshire, OX11 0QX, UK}
\affiliation{\mbox{Highly Correlated Electron Group, Physics Department, University of Johannesburg, P.O. Box 524,} Auckland Park 2006, South Africa}
\author{S.~Rayaprol}
\affiliation{UGC-DAE CSR, Mumbai Center, R-5 Shed, BARC, Trombay, Mumbai 400085, India}
\author{A.~D.~Hillier}
\affiliation{ISIS Neutron and Muon Facility, STFC, Rutherford Appleton Laboratory, Chilton, Oxfordshire, OX11 0QX, UK}
\author{J.~Sannigrahi}
\affiliation{ISIS Neutron and Muon Facility, STFC, Rutherford Appleton Laboratory, Chilton, Oxfordshire, OX11 0QX, UK}
\author{M.~Rotter}
\affiliation{McPhase Project, 01159 Dresden, Germany}
\author{M.~D.~Le}
\affiliation{ISIS Neutron and Muon Facility, STFC, Rutherford Appleton Laboratory, Chilton, Oxfordshire, OX11 0QX, UK}
\author{E.~V.~Sampathkumaran}
\affiliation{UGC-DAE CSR, Mumbai Center, R-5 Shed, BARC, Trombay, Mumbai 400085, India}
\affiliation{Homi Bhabha Centre for Science Education, Tata Institute of Fundamental Research, V. N. Purav Marg, Mankhurd, Mumbai, 400088 India.}

\date{\today}

\begin{abstract}
The spin-chain compound Sr$_3$NiPtO$_6$ is known to have a nonmagnetic ground state. We have investigated the nature of ground state of Sr$_3$NiPtO$_6$ using magnetic susceptibility $\chi(T)$, heat capacity $C_{\rm p}(T)$, muon spin relaxation ($\mu$SR) and inelastic neutron scattering (INS) measurements. The $\chi(T)$ and $C_{\rm p}(T)$ do not exhibit any pronounced anomaly that can be associated with a phase transition to a magnetically ordered state. Our $\mu$SR data confirm the absence of long-range magnetic ordering down to 0.04~K\@. Furthermore, the muon spin relaxation rate increases below 20~K and exhibits temperature independent behavior at low temperature, very similar to that observed in a quantum spin-liquid system. The INS data show a large excitation near 8~meV, and the analysis of the INS data reveals a singlet CEF ground state with a first excited CEF doublet state at $\Delta_{\rm CEF} = 7.7$~meV. The estimated CEF parameters reveal a strong planar anisotropy in the calculated $\chi(T)$, consistent with the reported behavior of the $\chi(T)$ of single crystal Sr$_3$NiPtO$_6$. We propose that the nonmagnetic singlet ground state and a large $\Delta_{\rm CEF}$ (much larger than the exchange interaction $\mathcal{J}_{\rm ex}$) are responsible for the absence of long-range magnetic ordering and can mimic a classical spin-liquid behavior in this quasi-1D spin chain system Sr$_3$NiPtO$_6$. The classical spin-liquid ground state observed in Sr$_3$NiPtO$_6$ is due to the single-ion property, which is different from the quantum spin-liquid ground state observed in geometrically frustrated  systems, where two-ion exchanges play an important role.
\end{abstract} 

\maketitle
\section{Introduction}

The triangular lattice systems have been of keen interests because of the novel exotic magnetic states that develop due to the presence of inherent geometric spin frustration  on these lattices. Of particular interests are the frustrated quasi-1D spin chain systems with general formula $A_3MM'$O$_6$ ($A$ denotes alkaline metals Sr, Ca, etc.\ and $M/M'$ denote transition metals), which exhibit a variety of unconventional magnetic properties due to strong spin-orbit coupling (SOC) and reduced dimensionality \cite{Nunez1997, Onnerud1996, Layland2000, Hillier2011, Yin2013, Liu2012, vajenine, Sampathkumaran2007, Sampathkumaran2002, Takeshita2006, Niitaka2001, Basu2014,Toth2016, Sampathkumaran2004, Sampathkumaran2004b, Takubo2005, Agrestini2011, Flahaut2003, Lefrancois2014, Mohapatra2007, Mikhailova2012,Nguyen1994, Niitaka1999, Nguyen1995, Ou2014, claridge1999, sarkar2010, Yin2013, Chattopadhyay2010, Zhang2010, McClarty2020, Niazi2001, Niazi2002, Birol2018, ONeal2019}. These spin chain materials crystallize in the K$_4$CdCl$_6$-type hexagonal structure (space group $R\bar{3}c$) consisting of 1D spin chains aligned along the {\it c}-axis which are formed by alternating face-sharing $M$O$_6$ trigonal prism and $M'$O$_6$ octahedra and are arranged on a triangular lattice in the {\it ab}-plane \cite{claridge1999, Nguyen1994,Nguyen1995}. As the presence of two different transition metal ions in these materials may provide multiple exchange paths, these materials are quite interesting from both experimental and theoretical view points. 

The Ca-based spin-chain compounds have been rigorously investigated \cite{Niitaka1999, Sampathkumaran2004, Takubo2005, Agrestini2011, Sampathkumaran2002, Sampathkumaran2004b, Sampathkumaran2007, Takeshita2006, Niitaka2001,Basu2014}, however detailed investigations of their Sr-based counterparts are still lacking. ${\rm Sr_3Mg}M{\rm O_6}$ ($M=$ Pt, Ir, Rh), ${\rm Sr_3ZnRhO_6}$, and Sr$_3M$IrO$_6$ ($M=$ Co, Cu, Ni, and Zn) are some of the known Sr-based spin-chain compounds. \cite{Nunez1997, Onnerud1996, Layland2000, Hillier2011, Yin2013, Liu2012, Birol2018, ONeal2019, Birol2018, Mikhailova2012, Mohapatra2007, Lefrancois2014, Flahaut2003, Toth2016, sarkar2010, Chattopadhyay2010, Ou2014, Zhang2010, McClarty2020, Niazi2001, Niazi2002}. Recently some of us performed microscopic investigations on Sr$_3$ZnRhO$_6$ \cite{Hillier2011}, Sr$_3$ZnIrO$_6$ \cite{McClarty2020} and Sr$_3$NiIrO$_6$ \cite{Toth2016} using neutron scattering and muon spin relaxation techniques. Sr$_3M$IrO$_6$ ($M=$ Co, Cu, Ni, and Zn) exhibits diverse physical properties due to the competing SOC, magnetic exchange, and crystal field interactions as a result of the coexistence of $3d$ and $5d$ metals \cite{Yin2013, Liu2012, Birol2018, ONeal2019, Birol2018, Mikhailova2012, Mohapatra2007, Lefrancois2014, Flahaut2003, Toth2016, sarkar2010, Chattopadhyay2010, Ou2014, Zhang2010, McClarty2020, Niazi2001, Niazi2002}. Further, in these Sr spin-chains, besides the presence of SOC, a strong intra-chain exchange coupling may also exist, resulting in an anisotropic exchange interaction as seen in Sr$_3$CuIrO$_6$ \cite{Yin2013}. The strong SOC may open a gap in the electronic spectrum driving the system to an insulating state, the so-called spin-orbit driven Mott insulator state. 

The spin-chain systems having both $M$ and $M'$ moment carrying transition metals present very interesting magnetic behavior, as is the case with Sr$_3$NiIrO$_6$ which hosts alternating chains of spin-1/2 (Ir$^{4+}$ occupying octahedral site) and spin-1 (Ni$^{2+}$ occupying trigonal prism site) ions along the $c$-axis  \cite{Mikhailova2012, Mohapatra2007, Lefrancois2014, Flahaut2003, Toth2016, sarkar2010, Chattopadhyay2010, Ou2014, Zhang2010, Birol2018, ONeal2019}. The magnetic state of Sr$_3$NiIrO$_6$ is characterised by a magnetic phase transition below 75~K and a spin-freezing type anomaly below 17~K in the magnetization, and a large single-ion anisotropy \cite{Mikhailova2012, Mohapatra2007, Lefrancois2014, Flahaut2003, Toth2016, sarkar2010, Chattopadhyay2010, Ou2014, Zhang2010, Birol2018, ONeal2019}. The neutron diffraction study finds magnetic peaks below 75~K, and the magnetic structure is described by propagation vector {\bf k} = (0, 0, 1) \cite{Lefrancois2014}. Based on the symmetry analysis an amplitude modulated antiferromagnetic arrangement of ferrimagnetic chains of Ir$^{4+}$ and Ni$^{2+}$ is suggested for the magnetic phase between 2~K and 75~K \cite{Lefrancois2014}. Despite the presence of a clear anomaly in the magnetization at 17~K, the neutron diffraction study does not see any change in the magnetic structure going through 17 K transition in Sr$_3$NiIrO$_6$. The inelastic neutron scattering (INS) study revealed a large energy gap ($\sim 30$~meV) in magnetic excitations with a quasi-1D nature of magnetic interaction \cite{Toth2016}. Infrared and optical spectroscopies reveal spin-charge-lattice entanglement and the presence of vibronic coupling in Sr$_3$NiIrO$_6$ \cite{ONeal2019}.

Sr$_3$NiPtO$_6$, which hosts magnetic Ni$^{2+}$ ($3d^8$, $S = 1$) and non-magnetic Pt$^{4+}$ ($5d^6$, $S=0$), is another interesting compound which is suggested to have a spin-liquid-like ground state \cite{Nguyen1994, vajenine, Mohapatra2007, Chattopadhyay2010, sarkar2010, claridge1999, Pandey2009, Pradipto2016, Narayana2009}. No evidence for a long-range magnetic ordering has been found for temperatures down to 1.8~K, though an anomaly associated with the low-dimensional short range magnetic ordering is seen in magnetic susceptibility around 25--30~K  \cite{Nguyen1994, Mohapatra2007}.  Magnetic measurements on single crystal Sr$_3$NiPtO$_6$ reveal a large easy-plane magnetic anisotropy which is described by a model of non-interacting Ni$^{2+}$ with a large magnetocrystalline anisotropy parameter $\mathcal{D}  \sim 7.5$--9.3~meV \cite{claridge1999, Chattopadhyay2010} indicating for a large-$\mathcal{D}$ magnetic phase in this compound.  A large-$\mathcal{D}$ magnetic phase is realized in quantum spin-chain systems when the condition $\mathcal{D}/\mathcal{J_{\rm ex}} > 1$ is fulfilled, where $\mathcal{J_{\rm ex}}$ represents the magnetic exchange interaction \cite{Chattopadhyay2010, Tasaki1991, Chen2003}. Theoretical attempts have been made to explain the observed magnetic behavior of this compound  \cite{sarkar2010, Pandey2009, Pradipto2016}. While density functional theory (DFT) calculations correctly predicts an easy-plane anisotropy, it underestimates the magnitude \cite{Pradipto2016}, whereas a wavefunction-based (CASPT2) approach yielded more accurate predictions \cite{Pradipto2016}.

Despite several experimental and theoretical works the nature of the magnetic ground state in Sr$_3$NiPtO$_6$ is still not clear and invites further investigations. The present work aims at exemplifying the nature of the magnetic ground state of Sr$_3$NiPtO$_6$ through microscopic investigations. We have probed the magnetic properties using muon spin relaxation ($\mu$SR) which finds an absence of long-range magnetic ordering down to 0.04~K. The inelastic neutron scattering measurements have enabled us to determine the crystal electric field (CEF) level scheme. The magnetization $M$ as a function of temperature $T$ and magnetic field $H$, and heat capacity $C_{\rm p}(T)$ data are also presented here to characterize the sample quality by comparing them with the reported results. Our remaining discussion is divided into three sections. We present experimental details  in Sec.~\ref{Exp}. The results of magnetization, heat capacity, $\mu$SR, and INS measurements are presented in Sec.~\ref{Result}. Finally a summary of results and conclusions is given in Sec.~\ref{Concl}.  Our results reflect that Sr$_3$NiPtO$_6$ indeed presents all the properties expected for a classical spin-liquid phase rather than a quantum spin-liquid behavior.

\section{Experimental Details}
\label{Exp}

A polycrystalline sample of Sr$_3$NiPtO$_6$ was prepared following the solid state reaction route in air starting with the high purity materials SrCO$_3$ (99.99\,\%), NiO (99.99\,\%) and PtO$_2$ (99.99\,\%) in powder form. The thoroughly mixed powders of the constituent components in stoichiometric ratio were calcined at 800 $^\circ$C for 24~h\@. After initial calcination, the powder was pressed into pelletized form and sintered at 1000 $^\circ$C for 9 days with three intermediate grindings as described in Ref.~\cite{Mohapatra2007}.

\begin{figure}%[b]
\includegraphics[width=\columnwidth]{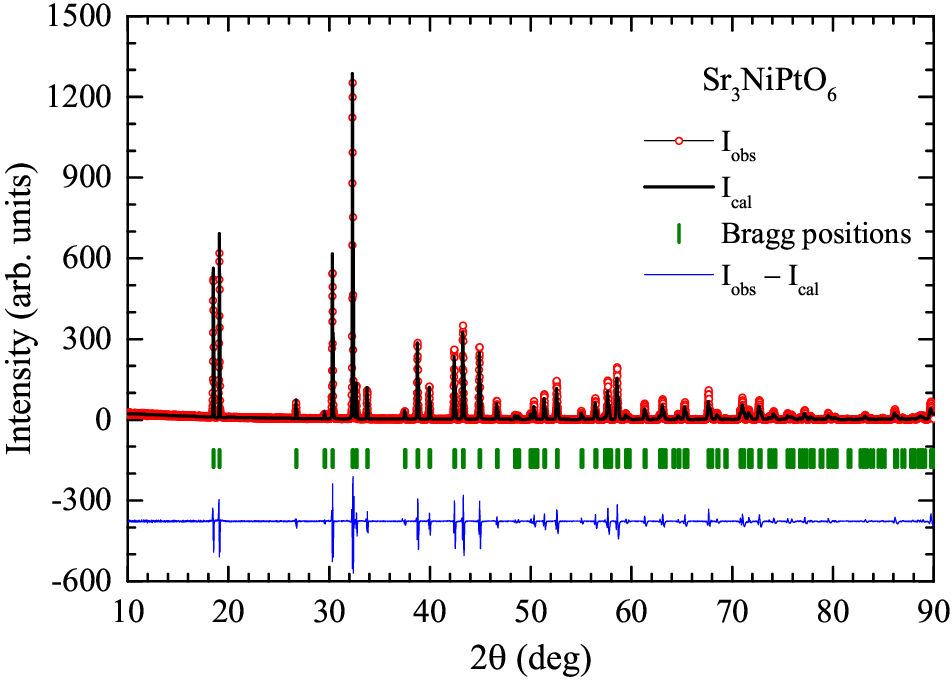}
\caption {X-ray powder diffraction pattern of Sr$_3$NiPtO$_6$ recorded at room temperature. The solid line through the experimental points is the Le Bail profile fit for K$_4$CdCl$_6$-type hexagonal structure (space group $R\bar{3}c$). The short vertical bars mark the Bragg peak positions. The lowermost curve represents the difference between the observed and calculated intensities.}
\label{XRD}
\end{figure}

The quality of the sample was examined by room temperature x-ray diffraction (XRD), performed on the powdered sample of Sr$_3$NiPtO$_6$, which is shown in Fig.~\ref{XRD}. 
The Le Bail profile fit for the reported hexagonal structure (space group $R\bar{3}c$) of Sr$_3$NiPtO$_6$ \cite{Nguyen1994, claridge1999} is shown in Fig.~\ref{XRD}. All the observed XRD peaks are consonant with the Bragg peaks expected in space group $R\bar{3}c$, and are well captured by the Le Bail profile fit. The single phase nature of the synthesized sample is evident from Fig.~\ref{XRD}. The lattice parameters $a=9.5888(1)$~{\AA} and $c=11.2003(2)$~{\AA} obtained from XRD are found to be in very good agreement with the reported values \cite{Nguyen1994, claridge1999}.

The magnetic susceptibility $\chi(T)$, isothermal magnetization $M(H)$, and heat capacity $C_{\rm p}(T)$ measurements were carried out using a Quantum Design SQUID VSM and physical properties measurement system (PPMS). The $\mu$SR experiments were carried out using the MuSR  spectrometer at the ISIS Neutron and Muon Facility, United Kingdom. $\mu$SR measurements were done in zero magnetic field. The powdered sample was mounted on an Ag-plate using GE-varnish. Temperatures down to 0.04~K were achieved by using a dilution stick in a $^4$He cryostat. For temperatures between 0.04~K and 4~K the data were collected under vaccuum, and for 4~K to 50~K He-exchange gas was added in the sample space \cite{isis}. Spin-polarized positive muons ($\mu^{+}$, mean lifetime of 2.2 $\mu$s, momentum of 28 MeV/c, and $\gamma_{\mu}$/2$\pi$ = 135.5 MHz T$^{-1}$) were implanted into the polycrystalline sample. The average spin polarization of the muons stopped within the sample is proportional to the decay positron asymmetry function $A(t)$ \cite{blundell, Hillier2022}. 

The inelastic neutron scattering measurements were carried out  on the  HET time-of-flight spectrometer at ISIS Neutron and Muon Facility with incident neutron energies ($E_i$) of 15~meV, 50~meV and 900~meV at several temperatures from 4.5~K to 300~K\@. The powdered sample (total mass of 8~g) was mounted in a thin Al-foil envelope, which was cooled down to 4.5~K using a closed-cycle refrigerator (CCR) in presence of He-exchange gas.

\section{Results}
\label{Result}

\subsection{Magnetic susceptibility and heat capacity}

\begin{figure}
\includegraphics[width=\columnwidth]{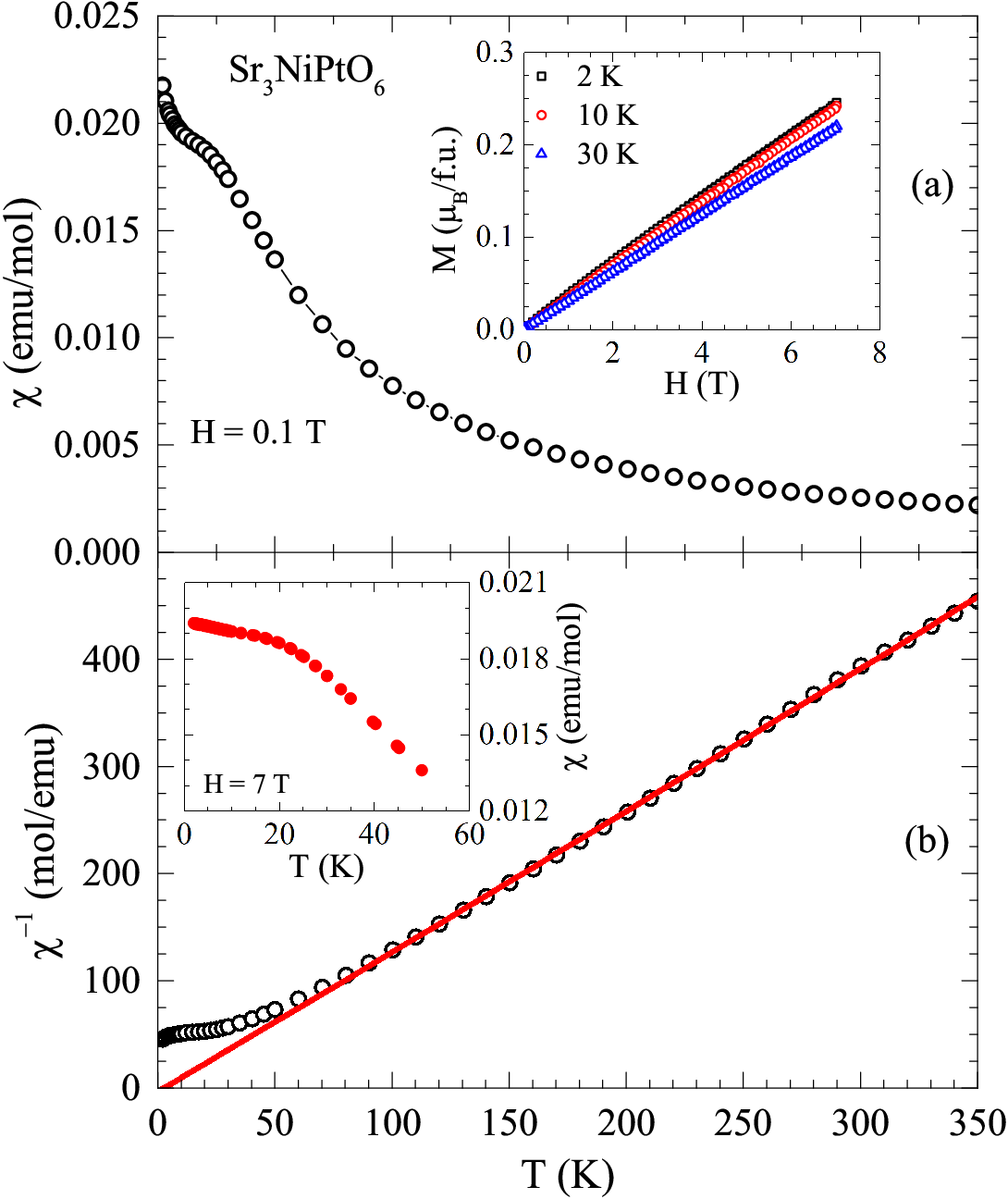}
\caption {(a) Temperature $T$ variation of magnetic  susceptibility $\chi$ of Sr$_3$NiPtO$_6$ for $ 1.8 \leq  T \leq 350$~K measured in a magnetic field $H = 0.1$ T\@. Inset shows the magnetization $M(H)$ isotherms at indicated temperatures.  (b) Inverse susceptibility plot $\chi^{-1}(T)$. The solid line is the fit to modified Curie-Weiss law. Inset shows the low-$T$ $\chi(T)$ measured in $H =7$~T\@.}
\label{MT}
\end{figure}

\begin{figure}
\includegraphics[width=\columnwidth]{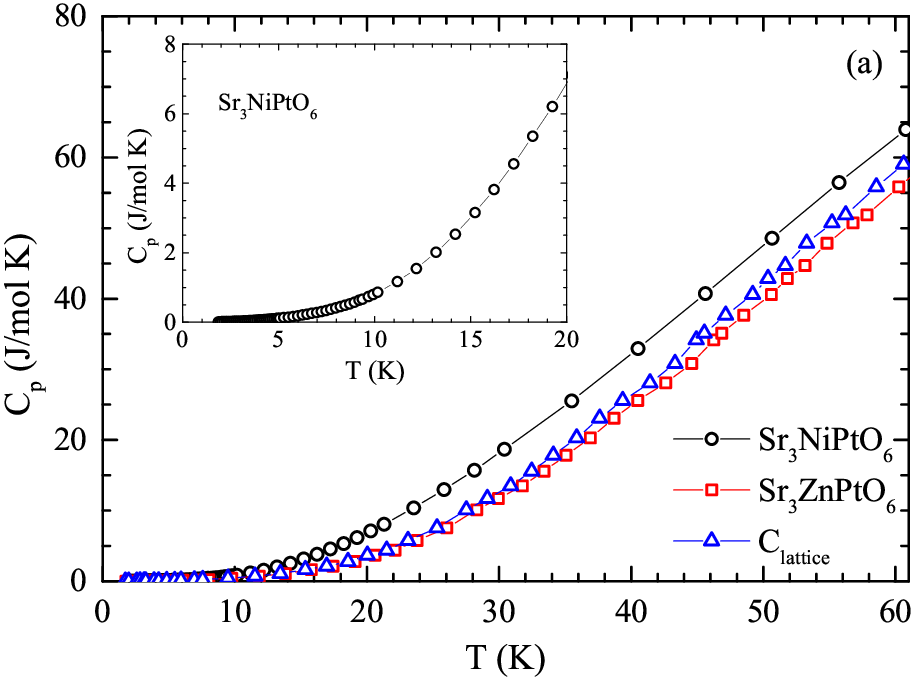}
\includegraphics[width=\columnwidth]{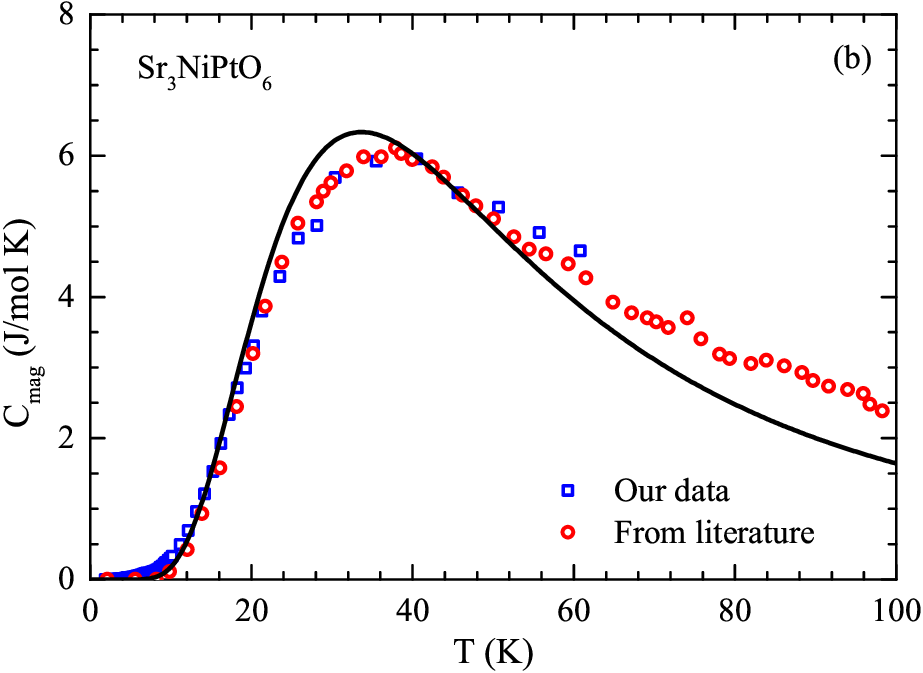}
\caption {(a) Temperature $T$ variation of the heat capcity $C_{\rm p}$ of Sr$_3$NiPtO$_6$ for $ 1.8 \leq  T \leq 60$~K measured in zero-field together with the $C_{\rm p}(T)$ of nonmagnetic reference compound Sr$_3$ZnPtO$_6$ from Ref.~\cite{Chattopadhyay2010}. The lattice contribution $C_{\rm lattice}(T)$ was obtained by scaling the $C_{\rm p}(T)$ of Sr$_3$ZnPtO$_6$ according to Eq.~(\ref{eq:Tscaling}) as discussed in text.  Inset: an expanded plot of $C_{\rm p}(T)$ of Sr$_3$NiPtO$_6$ for $T < 20$~K. (b) Magnetic heat capacity $C_{\rm mag}(T)$ for Sr$_3$NiPtO$_6$ along with the  $C_{\rm mag}(T)$ data from Ref.~\cite{Chattopadhyay2010} shown for comparison. The solid curve represents the $C_{\rm mag}(T)$ estimated using the CEF parameters obtained from the analysis of the inelastic neutron scattering data.}
\label{CP}
\end{figure}

The results of dc $\chi(T)$ and $M(H)$ measurements for Sr$_3$NiPtO$_6$ are shown in Fig.~\ref{MT}. We do not see any sharp anomaly associated with long-range magnetic ordering in $\chi(T)$ down to 1.8~K\@. The weak anomaly attributable to short range ordering, as reported for the polycrystalline samples \cite{Nguyen1994, Mohapatra2007}, is present in our $\chi(T)$ data. It is seen from Fig.~\ref{MT}(a) that at an applied magnetic field of 0.1~T the $\chi$ increases monotonically with decreasing temperature. At $T$ below $\sim 30$~K, a tendency to flatten is observed as often found in low dimensional systems. Additionally, an upturn is also present in our data below 10 K, known as a Curie-tail that could be due to the presence of paramagnetic impurities. The $\chi(T)$ follows the modified Curie-Weiss law, $\chi(T) = \chi_0 + C/(T-\theta_p)$, at temperatures above 100~K\@. Our fit of the inverse susceptibility to the modified Curie-Weiss law in the range $100~{\rm K} \leq T \leq 350$~K (see Fig.~\ref{MT}(b)) yielded $\chi_0 = 4.3(6) \times 10^{-5}$~emu/mol, the effective paramagnetic moment $\mu_{\rm eff} = 2.48(7) \,\mu_{\rm B}$ and a small but positive Weiss temperature $\theta_p =2.6(1.0)$~K\@.  The value of $\mu_{\rm eff}$ obtained from our data is close to that of the spin only value for Ni$^{2+}$ ($d^8$) which is 2.83\,$\mu_B$. Since Pt$^{4+}$ ($d^6$) is in the low spin ($S=0$) state  and is believed not to possess a magnetic moment, it is quite natural to achieve a $\mu_{\rm eff}$ close to the spin only value of Ni$^{2+}$. A small interchain interaction with a value of 0.10 meV was reported from the DFT calculation which turned out to be ferromagnetic \cite{sarkar2010} and can be correlated with the small, positive $\theta_p$ as observed from our data. The values of  $\theta_{\rm p}$ and $\mu_{\rm eff}$ obtained above differ from the previously reported values \cite{Nguyen1994, Mohapatra2007}. However, if we fit the $\chi(T)$ data by Curie-Weiss law we get a negative $\theta_{\rm p}$ ($=-8.5(6)$~K) and the value of  $\mu_{\rm eff}$ comes out to be $3.25(5)\,\mu_{\rm B}$ consistent with the literature values \cite{Nguyen1994, Mohapatra2007}.

Further, we measured $\chi(T)$ in presence of higher applied magnetic field of 7~T and we see that the Curie-tail becomes saturated and almost temperature independent at low-$T$ (see, inset in Fig.~\ref{MT}(b)). This Van-Vleck paramagnetism (the constant susceptibility) at low temperature could be related to a singlet CEF ground state. The isothermal $M(H)$ at 2~K, 10~K and 30~K are shown as an inset to  Fig.~\ref{MT}(a), and they are practically linear at all three temperatures for $H$ up to 7~T\@.

Figure~\ref{CP}(a) displays the $T$ variation of heat capacity of Sr$_3$NiPtO$_6$ between 1.8 to 60~K. For comparison, the $C_{\rm p}(T)$ data of nonmagnetic reference compound Sr$_3$ZnPtO$_6$ (from Ref.~\cite{Chattopadhyay2010}) is also presented.  The absence of an anomaly in $C_{\rm p}(T)$ data down to the lowest measured temperature clearly indicates an absence of long-range ordering. We see that the heat capacity is extremely small at temperatures below 5~K (see inset of Fig.~\ref{CP}(a)). Such a feature of heat capacity is an indication of the presence of a gap in the spin excitation spectra in Sr$_3$NiPtO$_6$.  

The $C_{\rm p}(T)$ data of Sr$_3$ZnPtO$_6$ \cite{Chattopadhyay2010} were used to estimate the phonon contribution to the heat capacity of Sr$_3$NiPtO$_6$. However, since the formula masses and unit cell volumes of Sr$_3$NiPtO$_6$ and Sr$_3$ZnPtO$_6$ are different, the measured $T$ of Sr$_3$ZnPtO$_6$ was scaled according to \cite{Anand2015a, Anand2015b}
\begin{equation}
\label{eq:Tscaling}
T^* = {T}\left(\frac{M_{\rm Sr_3NiPtO_6}}{M_{\rm Sr_3NiZnO_6}}\right)^{1/2}{\left(\frac{V_{\rm Sr_3NiPtO_6}}{V_{\rm Sr_3NiZnO_6}}\right)^{1/3}},
\end{equation}
where $M$ is the formula mass and $V$ the volume per formula unit. The scaled $C_{\rm p}(T^*)$ of Sr$_3$ZnPtO$_6$ was then taken as lattice contribution to separate out the magnetic contribution to the heat capacity $C_{\rm mag}(T) $ of Sr$_3$NiPtO$_6$. The $C_{\rm mag}(T) $ estimated this way is presented in Fig.~\ref{CP}(b) along with the literature data of $C_{\rm mag}(T) $ from Ref.~\cite{Chattopadhyay2010} and compared to the $C_{\rm mag}(T) $ obtained using the crystal electric field parameters based on the analysis of inelastic neutron scattering data (See Sec.~\ref{Sec:INS}). As can be seen from Fig.~\ref{CP}(b) our $C_{\rm mag}(T) $ agree very well with the literature data and present a broad Schottky type anomaly which is well captured by the CEF levels scheme. 

\begin{figure}
\includegraphics[width=\columnwidth]{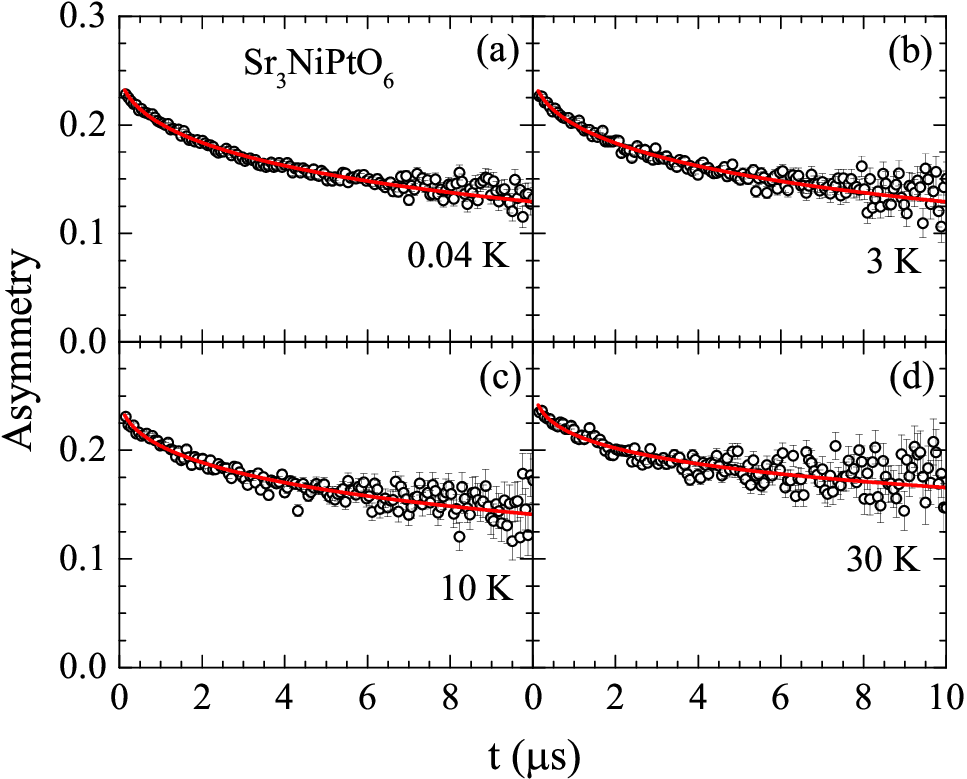}
\caption {Zero-field $\mu$SR asymmetry vs time $t$ spectra of Sr$_3$NiPtO$_6$ at indicated temperatures. The red solid curves are the fits to the $\mu$SR data by Eq.~(\ref{eq:exp}).}
\label{muSR1}
\end{figure}

\subsection{Muon Spin Relaxation}
The zero-field (ZF) asymmetry $\mu$SR spectra for Sr$_3$NiPtO$_6$ are shown in Fig.~\ref{muSR1} for four representative temperatures 0.04~K, 3~K, 10~K and 30~K\@. As can be seen from Fig.~\ref{muSR1}, down to 0.04~K ZF signals do not show any clear sign of oscillations within the analyzed time window up to 20 $\mu$s. The observation that at both high and low temperatures (down to 0.04~K) $\mu$SR spectra relax in a similar way with nearly same values of the initial asymmetry excludes the presence of long-range magnetic ordering in Sr$_3$NiPtO$_6$. The zero field spectra also lack the recovery of the one-third of the polarization which signifies the absence of static random field, which gives a Kubo-Toyabe type relaxation. We thus see that the $\mu$SR spectra do not show any evidence for a magnetic phase transition down to 0.04~K\@.

\begin{figure}
\includegraphics[width=\columnwidth]{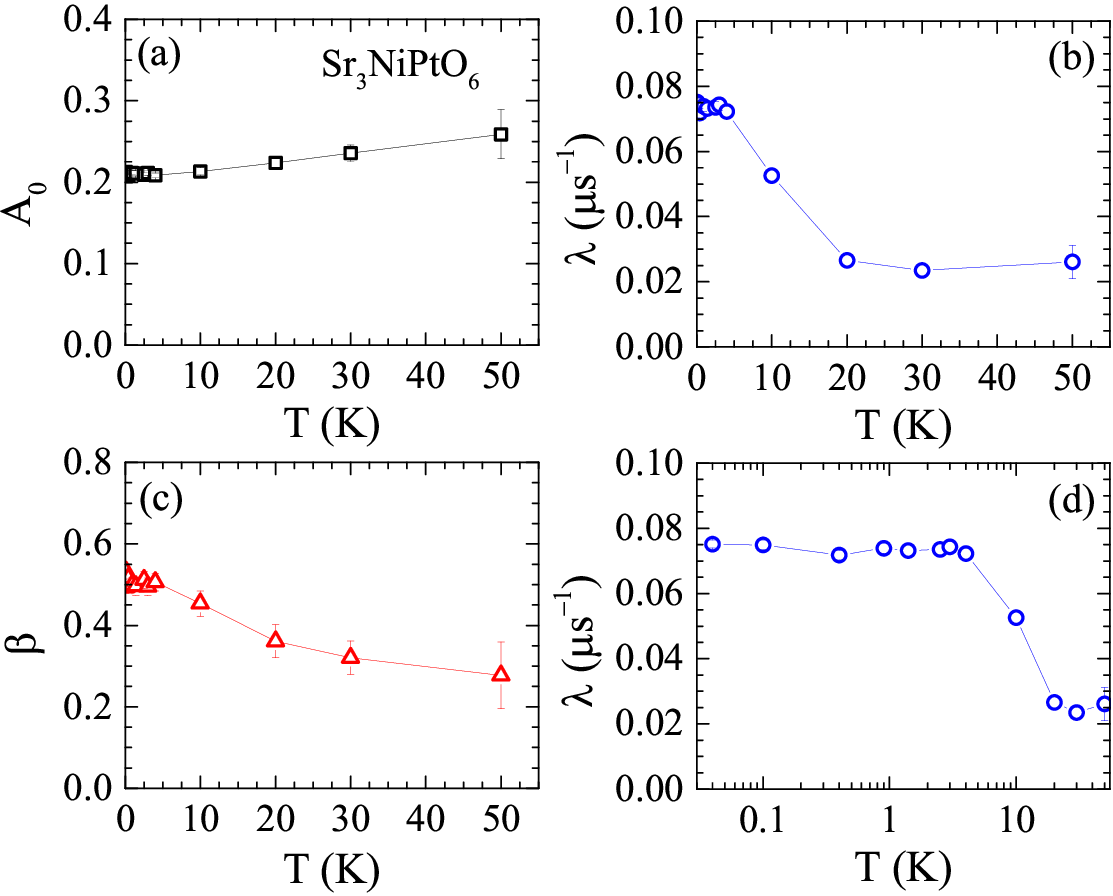}
\caption {Temperature $T$ dependence of the fit parameters: (a) muon initial asymmetry $A_0$, (b) relaxation rate $\lambda$, and (c) exponent $\beta$ obtained from the fitting of the zero-field $\mu$SR spectra by Eq.~(\ref{eq:exp}). (d) The $T$ dependence of $\lambda$ plotted on semi-logarithmic scale.}
\label{muSR2}
\end{figure}

\begin{figure*}
\includegraphics[width=\textwidth]{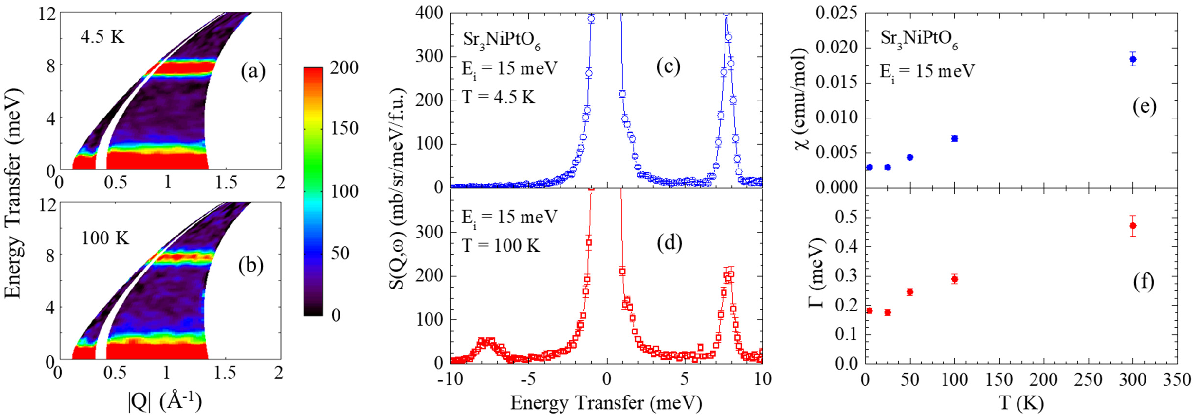}
\caption {Inelastic neutron scattering response, a color-coded contour map of the intensity, energy $E$ transfer vs momentum $|{Q}|$ transfer for Sr$_3$NiPtO$_6$ measured at representative temperatures (a) 4.5~K, and (b) 100~K, using neutrons with incident energy $E_i=15$~meV.  Corresponding $Q$-integrated (between 0 to 2 \AA$^{-1}$) one-dimensional (1D) energy cuts of INS responses, inelastic scattering intensity $S(Q,\omega)$ versus $E$ at (c) 4.5~K, and (d) 100~K\@. Temperature $T$ dependence of the fit parameters, (e) susceptibility $\chi$ (amplitude of the Lorentzian) vs $T$, and (f) linewidth $\Gamma$ vs $T$, obtained from fitting the 1D energy cuts to a Lorentzian function convoluted with the instrument resolution}
\label{INS1}
\end{figure*}

The form of the relaxation depends on the distribution and time dependence of the local magnetic fields around the muon implanted sites. Our ZF data are well fitted by a stretched exponential function,
\begin{equation}
A = A_0\exp[-({\lambda}t)^\beta]+ A_{\rm BG},
\label{eq:exp}
\end{equation} 
where $A_0$ is the initial asymmetry, $\lambda$ is the $\mu^+$ spin relaxation rate, $\beta$ is an exponent and $A_{\rm BG}$ is the background contribution to muon asymmetry. We choose constant background $A_{\rm BG} \approx 0.04$ representing muons that missed the sample and stopped on the Ag sample holder. The fitting parameters obtained from the fits of ZF-$\mu$SR spectra by Eq.~(\ref{eq:exp}) are presented in Fig.~\ref{muSR2}. Fits are shown by red solid curves in Fig.~\ref{muSR1}. The asymmetry $A_0$ shows a weak temperature dependence, decreases from 0.26 (at 50~K) to 0.21 (at 4~K), and remains unchanged between 0.04~K and 4~K\@. The relaxation rate $\lambda$ shows a strong temperature dependence, increases rapidly below 20~K and eventually saturates at $T\leq  4$~K as can be seen more clearly from the logarithmic scale plot in Fig.~\ref{muSR2}(d). The exponent $\beta$ is also seen to increase with decreasing $T$ [Fig.~\ref{muSR2}(c)]. 
The fact that the $\mu$SR data could not be described by the classical exponential function ($\beta  =1$) reflects that our system is not a simple paramagnetic system. The stretched exponential fit yields $\beta$ values much lower than one, which reveals that there is a distribution of relaxation times in Sr$_3$NiPtO$_6$ as expected when different spins do fluctuate on different time scales. For a spin-glass system $\beta$ is expected to reach a value of 1/3 when the system approaches to the spin-glass freezing temperature. As can be seen from Fig.~\ref{muSR2}(c), for the present compound $\beta$ reaches to 1/2 which is not consistent with the spin-glass freezing in this compound. However, all these are consistent with a spin-liquid like behavior in Sr$_3$NiPtO$_6$. Further, the observed temperature dependence of $\lambda$ ($T$-independent behavior of $\lambda$ at low-$T$) is very similar to that observed in frustrated magnet $\rm{SrCr_8Ga_4O_{19}}$ \cite{Uemura1994}, and in quantum spin-liquid materials NaYbS$_2$ \cite{Sarkar2019} and YbMgGaO$_4$ \cite{Li2016}. This signifies a fluctuating spin-dynamics consistent with the proposed spin-liquid like behavior in Sr$_3$NiPtO$_6$. We would like to point out that the quantum spin-liquid behavior (no magnetic ordering down to 0~K) arises due to magnetic frustration (two or more ions exchange interactions), while the classical spin-liquid we have observed in Sr$_3$NiPtO$_6$, is due to a single ion effect arising from the non-magnetic ground state.

\begin{figure}
\includegraphics[width=7cm]{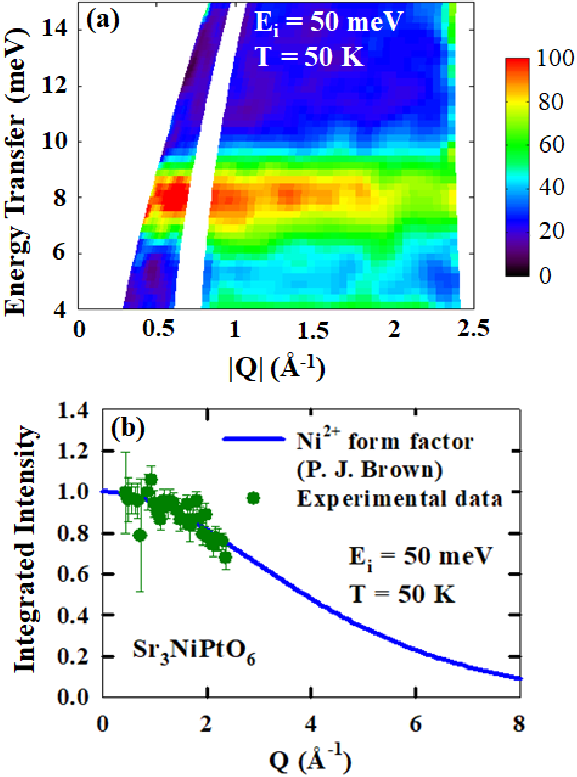}
\caption {(a) Color-coded inelastic neutron scattering (INS) intensity of Sr$_3$NiPtO$_6$ measured at 50~K with neutrons of incident energy $E_i= 50$~meV. (b) Energy integrated, $Q$-dependent intensity between 6.5 and 9.5~meV  (filled symbols) obtained from the INS response in (a). The solid curve in (b) shows the squared magnetic form factor [$F^{2}$(Q)] of Ni$^{2+}$ from P. J. Brown \cite{Brown} scaled to 1 at $Q=0$.}
\label{INS2}
\end{figure}

\begin{figure}
\includegraphics[width=3in]{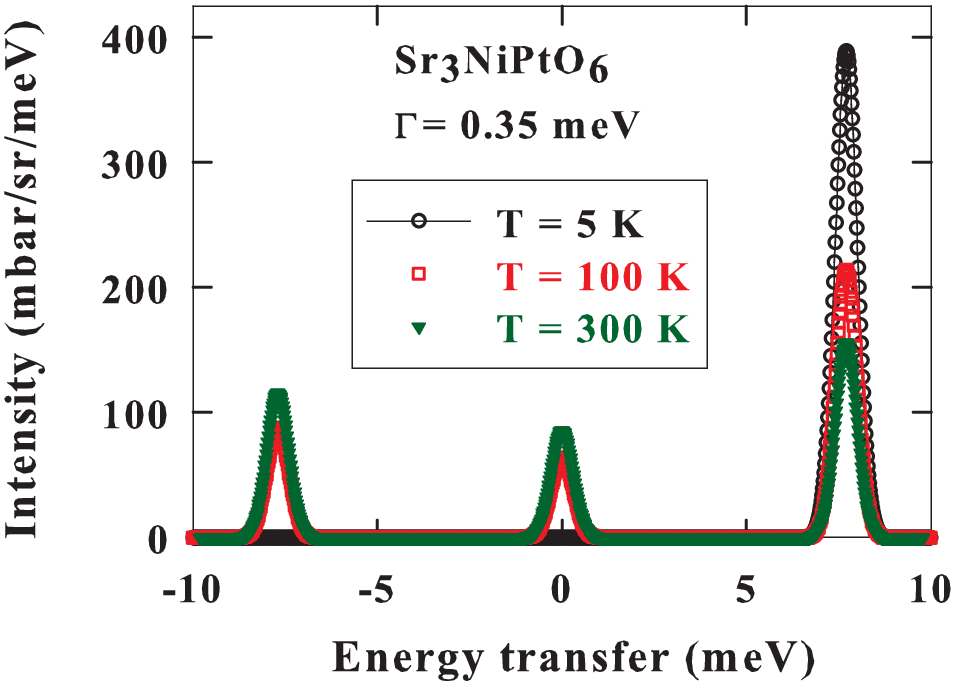}
\caption {Calculated crystal electric field excitations of Ni$^{2+}$ at 5~K, 100~K and 300~K for Sr$_3$NiPtO$_6$ calculated using McPhase program for the $L_{lm}$ parameters obtained from the analysis of the inelastic neutron scattering data. The excitations are calculated at $Q=0$.}
\label{fig:CEFcal1}
\end{figure}

\begin{figure}
\includegraphics[width=\columnwidth]{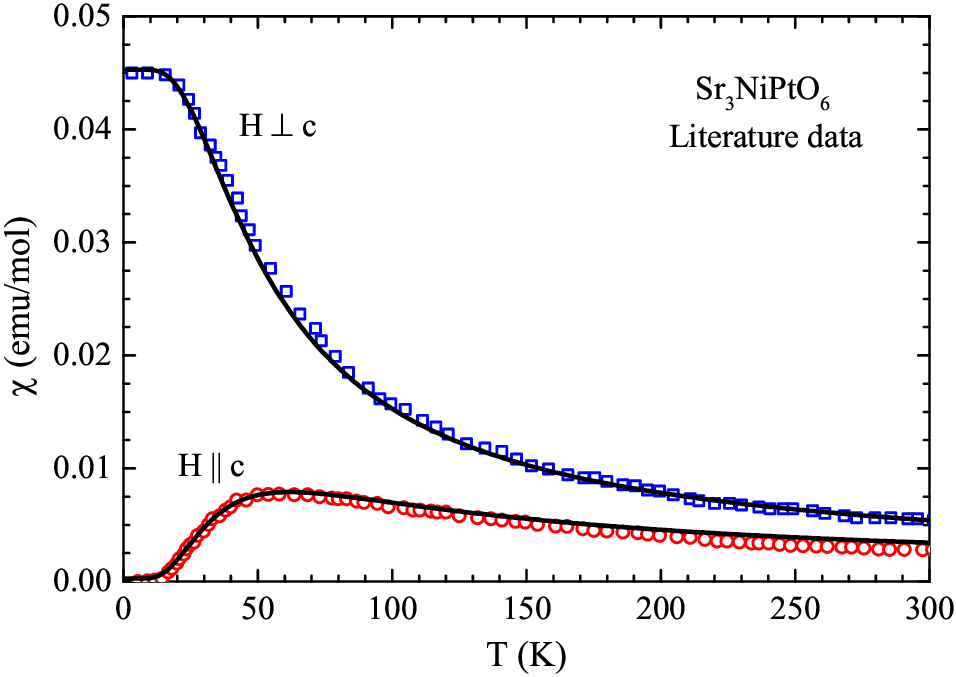}
\caption {Calculated single crystal magnetic susceptibility $\chi(T)$ shown by the solid lines and the experimental data from Ref. ~\cite{Chattopadhyay2010} shown by symbols for comparison. The $\chi(T)$ was calculated using the CEF parameters obtained from the analysis of the inelastic neutron scattering data.}  
\label{fig:CEFcal2}
\end{figure}

\subsection{\label{Sec:INS} Inelastic Neutron Scattering} 

Figure~\ref{INS1}(a-b) shows the temperature dependent two dimensional (2D) color-coded intensity maps, energy $E$ transfer vs momentum $|Q|$ transfer for Sr$_3$NiPtO$_6$ measured with incident neutrons of energy $E_i=15$~meV. The data are from the low-angle scattering banks of the HET spectrometer between 3.5$^\circ$ and 30$^\circ$. At 4.5~K we see very strong excitation near 8~meV. The intensity of this excitation was absent in the high angle bank at 110$^\circ$ and 135$^\circ$ which reflects the magnetic origin of this excitation. Further, with increasing temperature the intensity of this excitation decreases, the presence of the excitation could be clearly seen even at 300~K (data not shown). $Q$-integrated one-dimensional (1D) energy cuts of INS responses, inelastic scattering intensity $S(Q,\omega)$ versus $E$ are plotted in Fig.~\ref{INS1}(c-d) for representative temperatures 4.5~K and 100~K\@. The decrease in the intensity of the excitation with increasing temperature can be seen very clearly in the 1D cuts. Further, we see that with increasing temperature the inelastic peak on the neutron energy gain side becomes visible, which is due to the thermal population factor of the 8~meV excitation.  It is to be noted that the temperature independent  shoulder on the elastic line near 1~meV energy transfer is the background from the CCR due to multiple scattering. 

Figure~\ref{INS2}(a) shows the color-coded 2D-intensity map of the inelastic neutron scattering response for the measurement with a higher incident energy $E_i=50$~meV at 50 K in order to cover the larger range of $E$-$Q$ space to investigate the $Q$ dependence of the intensity of the 8 meV excitation. In Fig.~\ref{INS2}(b) we have plotted the $Q$ dependence of the energy integrated  intensity (integrated between 6.5 and 9.5~meV) of the 8~meV excitation. It is evident that the intensity of this peak decreases with increasing $Q$, which indicates the magnetic nature of this excitation. To further confirm that the intensity of this peak follows the Ni$^{2+}$ magnetic form factor squared $F^{2}(Q)$ \cite{Brown}, we have also plotted the squared magnetic form factor, the solid curve in Fig.~\ref{INS2}(b) which passes through the experimental data. This confirms that the 8~meV peak originates from the Ni$^{2+}$ ion.  We also performed high energy $E=900$~meV measurements at 5~K and 300~K and we did not find any clear signature of the presence of any additional magnetic excitation in this energy range (see Fig.~\ref{INS3} in Appendix).

In order to gain further information on the nature of 8~meV excitation, we have fitted the $S(Q,\omega)$ versus $E$ data using a Lorentzian function convoluted with the instrument resolution function which provides information about the static susceptibility, position and linewidth of this peak. The static susceptibility (amplitude of Lorentzian) and linewidth  $\Gamma$ (HWHM) obtained from the fits of $S(Q,\omega)$ are plotted in Figs.~\ref{INS1}(e-f). Further, to check the temperature dependence of linewidth $\Gamma$ we also plotted $\Gamma$ vs $T^{1/2}$ as well as $\Gamma$ vs $T^{2}$ (figure not shown), and we found that overall $T$ dependence of $\Gamma$ between 4.5~K and 300~K is described best by a linear behaviour. 

We estimate the effective paramagnetic moment from the 8~meV peak by using the moment sum rule, according to which
\begin{equation}
\int \frac{S(Q,\omega)}{F^{2}(Q)}d\omega = A \mu_{\rm eff}^2
\end{equation}
where the constant $ A=(2/3)(r_0/2)^2$ with $r_0=g_{\rm n} r_{\rm e}/2$, $r_{\rm e}$ being the classical electron radius and $g_{\rm n}$ the neutron $g$-factor. With the values of $g_{\rm n}$ and $r_{\rm e}$, $A = 48.43$~mb/sr/$\mu_{\rm B}^2$. The integrated intensity for the 8~meV peak is found to be 350(7)~mb/sr; accordingly we obtain  $\mu_{\rm eff}  = 2.68(4)\, \mu_{\rm B}$ at 4.5~K which is in good agreement with the value of $\mu_{\rm eff} = 2.5 \,\mu_B$  obtained from the analysis of $\chi(T)$ and with the theoretically expected value of $2.83 \,\mu_B$ for Ni$^{2+}$ ions. As the deduced $\mu_{\rm eff}$ is close to the full value expected for Ni$^{2+}$, it implies that all the spectral weight has been accounted for and no other excitation should be visible.

We now present the analysis of CEF excitations based on the CEF model. The crystal field Hamiltonian (in Wybourne normalization) relevant for the point symmetry $D_3$ of Ni$^{2+}$ ions in Sr$_3$NiPtO$_6$ is given by
\begin{equation}
H_{\rm CEF} = L_{20} \hat{T}_{20}+ L_{40} \hat{T}_{40}+ L_{43} \hat{T}_{43},
\end{equation}
where $L_{lm}$ are the crystal-field parameters with he same normalisation as the Wybourne parameters, but are real numbers,  and $\hat{T}_{lm}$ are tensor operators defined in the McPhase program ~\cite{McPhase}, which are the Hermitian combinations of the Wybourne tensor operators $\hat{C}_{lm}$. The relation between CEF parameters $L_{lm}$ used here and conventional Stevens CEF parameters $B_n^m$ \cite{Stevens} is given by $B_n^m = \lambda_{lm} L_{lm} \langle L || \theta_l || L \rangle$, where the values of  normalization constants $\lambda_{lm}$ are given in~\cite{McPhase} and $\langle L || \theta_l || L \rangle$ are the orbital Stevens operator equivalent factors ($\theta_l$ for $l=0$, 2, 4, 6) given for Ni$^{2+}$ in Abragam and Bleaney ~\cite{Abragam and Bleaney}.

We used the  McPhase program ~\cite{McPhase} for analysis of our inelastic data and calculation of the physical properties. First we use the point-charge model for an initial estimation of the crystal field parameters ($L_{lm}$). The estimated L$_{lm}$ CEF parameters from the point charge model gave an INS excitation  near 15~meV, as compared to the observed 8~meV excitation. Then we kept the value of $L_{40}$ and $L_{43}$ fixed from the point charge estimation and manually vary the value of $L_{20}$ to get good agreement between the observed and calculated CEF excitations.  We found that when we divide the point charge value of $L_{20}$ by a factor 3.5 we get very good agreement between the data and calculation. The ground state is a singlet with a first excited doublet at 7.7~meV (two singlets having same energy), and the overall splitting energy is found to be $\approx 605$~meV. The $L_{lm}$ parameters obtained are (in meV): $L_{20} = 80.26$, $L_{40} =-609.96$ and $L_{43} = -180.15$ (corresponding value of Stevens parameters are: $B_2^0=0.764$, $B_4^0=- 0.484$ and $B_4^3 =- 3.383$ in meV). All 21 energy levels of Ni$^{2+}$ ($L=3$ and $S=1$) obtained from the analysis of the INS data are listed in Table~\ref{tab:CEF}. The ground state wave function is found to be
\begin{equation}
\begin{split}
\Psi_0  =  -0.885|^3F_4,m_J=0\rangle + 0.252|^3F_2,m_J=0\rangle  \\ 
   + 0.198|^3F_3,m_J=3\rangle + 0.198|^3F_3,m_J=-3\rangle, \nonumber
 \end{split}
\end{equation}
 and, the wave function for the first excited state is 
\begin{equation}
\begin{split}
\Psi_1=        \pm 0.706|^3F_4,m_J=\mp 1 \rangle    -0.530|^3F_4,m_J=\mp 4 \rangle  \\
 -0.417|^3F_3,m_J=\mp 1 \rangle   \pm0.129|^3F_2,m_J=\mp 1 \rangle.  \nonumber
 \end{split}
\end{equation}

\begin{table}
\caption{\label{tab:CEF} Crystal electric field level scheme obtained from the analysis of the inelastic neutron scattering data. The 21-fold degenerate ground multiplet of Ni$^{2+}$ ($L=3$ and $S=1$) splits into a combination of 7 singlets and 7 doublets. Splitting energies are relative to the ground state energy (set to zero).}
\begin{ruledtabular}
\begin{tabular}{lcc}
Level & Splitting Energy (meV) & Splitting Energy (K) \\
\hline
Singlet & 0 & 0\\
Doublet & 7.7  & 89.4\\
Doublet &  134.9 & 1565.6  \\
Doublet &  192.5 & 2233.8 \\
Singlet & 202.8 & 2352. 9 \\
Singlet & 231.5 & 2686.3 \\
Doublet & 280.3 & 3253.2 \\
Singlet & 300.6 & 3488.0\\
Doublet & 345.6 & 4011.0\\
Singlet & 359.6 & 4172.9\\
Singlet & 451.7 & 5241.3 \\
Singlet & 463.6& 5380.2 \\
Doublet &  552.4 & 6409.9\\
Doublet &  605.0 & 7020.5 \\
\end{tabular}
\end{ruledtabular}
\end{table}

The calculated CEF excitations for Ni$^{2+}$ using McPhase for the above obtained $L_{lm}$ parameters are shown in Fig.~\ref{fig:CEFcal1} for temperatures 5~K, 100~K and 300~K\@. We can see that the 8~meV excitation is well reproduced by the CEF parameters. It is to be noted that only one strong CEF excitation near 8~meV is seen in the calculation, which is in agreement with the experimental observation. Next we use the CEF parameters to calculate the single crystal magnetic susceptibility $\chi(T)$ for $H \parallel c$ and $H \perp c$ which is presented in Fig.~\ref{fig:CEFcal2}. Our calculated $\chi(T)$ data clearly show the strong anisotropy and are in qualitative agreement with the experimental $\chi(T)$ of single crystal Sr$_3$NiPtO$_6$ reported in Ref.~\cite{Chattopadhyay2010}, which are also presented in Fig.~\ref{fig:CEFcal2} for comparison. We also calculated the magnetic heat capacity $C_{\rm mag}(T)$ using the CEF parameters which is presented in Fig.~\ref{CP}(b) along with our experimental $C_{\rm mag}(T)$ data and the literature data from Ref.~\cite{Chattopadhyay2010}. Our CEF model heat capacity reproduces the experimental $C_{\rm mag}(T)$ data of Sr$_3$NiPtO$_6$ very well. 

The occurrence of long-range magnetic ordering in a singlet ground state system depends on the relative strength of the exchange interaction ${\mathcal{J}_{\rm ex}}$ and the CEF splitting energy $\Delta_{\rm CEF}$ (between the ground state and the first excited state). An induced moment ordering is possible provided the critical value of ${\mathcal{J}_{\rm ex}}$ fulfills the condition $4 {\mathcal{J}_{\rm ex}} \alpha^2/\Delta_{\rm CEF} \geq 1$, where $\alpha$ is the matrix element of angular momentum between the CEF ground state and CEF excited state \cite{Trammel1963, Cooper1967, Wang1968}. The exchange energy ${\mathcal{J}_{ex}}$ can be estimated from the value of $\theta_{\rm p}$, as \cite{Johnston2012, Johnston2015}
\begin{equation}
\theta_{\rm p} = -\frac {\mathcal{J}_{\rm ex}\, S(S+1)}{3 k_{\rm B} },
\label{eq:Jex}
\end{equation}
where $\mathcal{J}_{\rm ex} =  \sum_{j}\mathcal{J}_{ij} $, and $k_{\rm B}$ is Boltzmann's constant. For Sr$_3$NiPtO$_6$, the analysis of $\chi(T)$ has yielded $\theta_{\rm p} = +2.6$~K, accordingly for Ni$^{2+}$ ($3d^8$, $S = 1$), $\mathcal{J}_{\rm ex} = -0.336$~meV. This value of $|\mathcal{J}_{\rm ex}| = 0.336$~meV ($\approx 3.9$~K) is much smaller than the CEF splitting energy $\Delta_{\rm CEF}= 7.7$~meV ($\approx 89$~K) between the ground state singlet and the first excited doublet, and does not fulfill the condition for induced moment ordering.  It appears that it is the presence of the nonmagnetic CEF-split singlet ground state and $\Delta_{\rm CEF} \gg \mathcal{J}_{\rm ex}$ which together create a very unfavorable condition for the development of long-range magnetic ordering, and in the absence of long-range ordering a classical spin-liquid behavior (originating from single-ion exchange effect) is observed in Sr$_3$NiPtO$_6$.

\section{Conclusions} 
\label{Concl}
We have investigated the electronic and magnetic ground state of Sr$_3$NiPtO$_6$ using $\chi(T)$, $M(H)$, $C_{\rm p}(T)$, muon spin relaxation and inelastic neutron scattering measurements in order to understand the proposed spin-liquid behavior in this quasi-one dimensional spin chain system.  $\chi(T)$ and $C_{\rm p}(T)$  do not show any sign of long-range magnetic ordering down to 2~K\@. The absence of long-range ordering down to 0.04~K is further confirmed by the $\mu$SR data, consistent with the spin-liquid behavior.  The $\mu$SR data are well described by a stretched exponential relaxation function. Our analysis of zero-field $\mu$SR spectra reveals a weak relaxation above 20 K, and below 20~K there is an increase in relaxation rate with a nearly temperature independent saturating behavior between 4~K and 0.04~K\@. The $\mu$SR data reflect a spin-liquid like spin dynamics in Sr$_3$NiPtO$_6$. 

The inelastic neutron scattering finds a strong excitation near 8~meV. The analysis of INS data reveals a singlet CEF ground state with a first excited doublet state at $\Delta_{\rm CEF} = 7.7$~meV. The magnetic susceptibility calculated using the CEF parameters obtained from the analysis of INS data reveals a strong planar anisotropy, and is qualitatively in very good agreement with the reported single crystal susceptibility data. The absence of long-range magnetic ordering, and hence the spin-liquid like behavior in Sr$_3$NiPtO$_6$ can be attributed to the presence of a nonmagnetic CEF singlet ground state and a large CEF splitting between the ground state and the first excited doublet, $\Delta_{\rm CEF} \gg  \mathcal{J}_{\rm ex}$. Further it is the single-ion exchange property of Ni$^{2+}$ which is responsible for the spin-liquid like behavior in Sr$_3$NiPtO$_6$. This mechanism is different from the two-ion exchanges driven quantum spin-liquid state observed in geometrically frustrated systems \cite{Rau2019, Broholm2020,Anderson1973}, and hence we call it to be a classical spin-liquid behavior in Sr$_3$NiPtO$_6$. Our results will pave the way to understand the classical spin-liquid behavior in many singlet ground state systems based on transition metals, rare-earths, and actinides compounds.

\section*{ACKNOWLEDGEMENT}
DTA would like to thank the Royal Society of London for the Newton Advanced fellowship funding between UK and China, and International Exchange funding between UK and Japan. DTA also thanks EPSRC UK for funding (Grant Ref.EP/W00562X/1) J.S. would like to thank the European Union’s Horizon 2020 research and innovation programme under the Marie Skłodowska-Curie grant agreement (GA) No 665593 awarded to the Science and Technology Facilities Council. E.V.S. thanks Department of Atomic Energy, Government of India, for awarding Raja Ramanna Fellowship (Order No. 1003/4/2021/RRF/R\&D-II/8335, dated 22.07.2021). We thank the ISIS facility for providing the beam time, and the muon and neutron data are available upon request.

\section*{Appendix: INS data measured with $E_i = 900$~meV}

\begin{figure}%[h]
\includegraphics[width=\columnwidth]{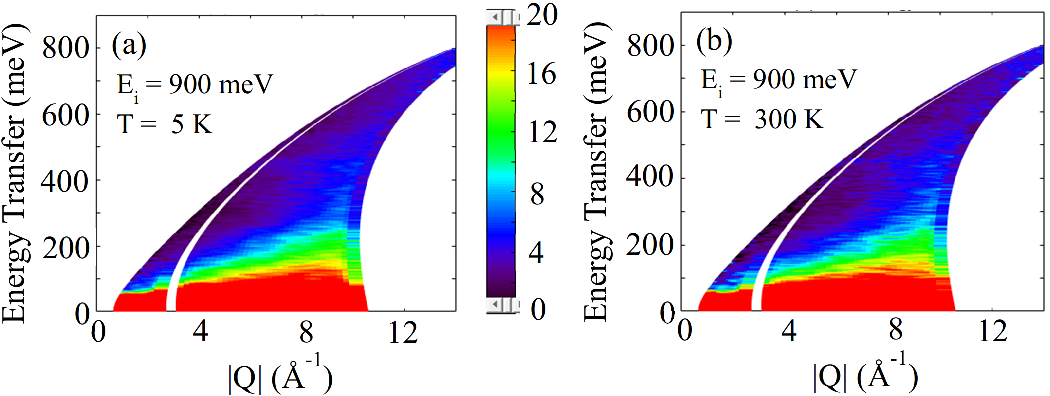}
\includegraphics[width=\columnwidth]{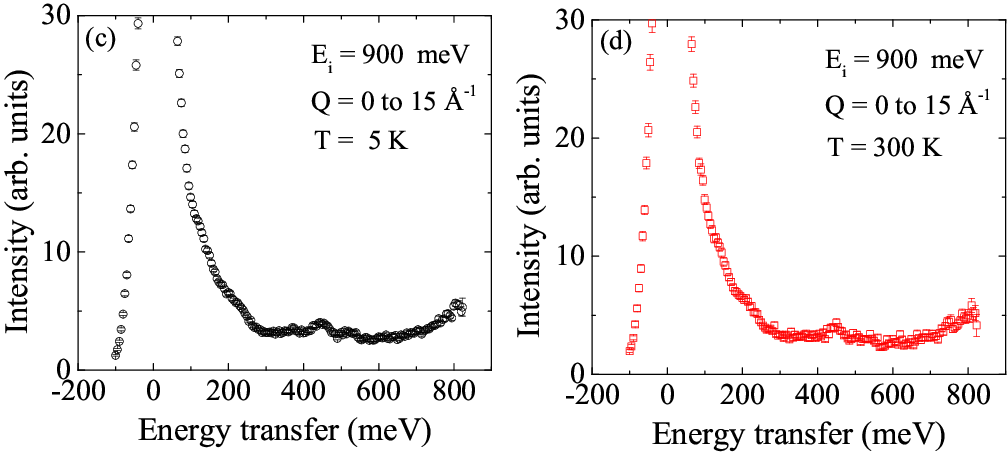}
\caption { Color-coded inelastic neutron scattering (INS) intensity map of the energy $E$ transfer vs momentum $|Q|$ transfer for Sr$_3$NiPtO$_6$ measured at (a) 5~K and (b) 300~K with neutrons of incident energy $E_i= 900$~meV. $|Q|$-integrated (between 0 and 15 \AA$^{-1}$) one dimensional energy cuts of INS responses at (c) 5~K and (d) 300~K\@.}
\label{INS3}
\end{figure}

Figure~\ref{INS3} shows the color-coded intensity maps, $E$ transfer vs $|Q|$ transfer for Sr$_3$NiPtO$_6$ measured at 5~K and 300~K with $E_i=900$~meV along with the plots of $Q$-integrated one-dimensional energy cuts.

\end{document}